\documentclass{aa}  

\usepackage{graphicx}
\usepackage{float}
\usepackage{natbib}
\bibpunct{(}{)}{;}{a}{}{,}
\usepackage{txfonts}
\usepackage{color}

\begin{document}

   \title{Toward the measurement of neutrino masses: Performance of cosmic magnification with submillimeter galaxies}
   \titlerunning{Massive neutrinos and the }
   \authorrunning{Cueli M. M. et al.}

    \author{Cueli M. M.\inst{1,2}, Cabo, S. R.\inst{3,4}, Gonz{\'a}lez-Nuevo J.\inst{3,4}, Bonavera L.\inst{3,4}, Lapi A.\inst{1,2,5,6}, \\Viel, M.\inst{1,2,6,7}, Crespo D.\inst{3,4}, Casas J. M.\inst{3,4} and Fernández-Fernández, R.\inst{3,4}}

  \institute{
  $^1$SISSA, Via Bonomea 265, 34136 Trieste, Italy\\
    $^2$IFPU - Institute for fundamental physics of the Universe, Via Beirut 2, 34014 Trieste, Italy\\
    $^3$Departamento de Fisica, Universidad de Oviedo, C. Federico Garcia Lorca 18, 33007 Oviedo, Spain\\
             $^4$Instituto Universitario de Ciencias y Tecnologías Espaciales de Asturias (ICTEA), C. Independencia 13, 33004 Oviedo, Spain\\
$^5$IRA-INAF, Via Gobetti 101, 40129 Bologna, Italy\\
$^6$INFN-Sezione di Trieste, via Valerio 2, 34127 Trieste,  Italy\\
$^7$INAF-OATS, Via G. B. Tiepolo 11, I-34131 Trieste, Italy}

   \date{}

 
  \abstract
   {The phenomenon of magnification bias can induce a non-negligible angular correlation between two samples of galaxies with nonoverlapping redshift distributions. This signal is particularly clear when background submillimeter galaxies are used, and has been shown to constitute an independent cosmological probe.}
   {This work extends prior studies on the submillimeter galaxy magnification bias to the massive neutrino scenario, with the aim being to assess its sensitivity as a cosmological observable to the sum of neutrino masses.}
   {The measurements of the angular cross-correlation function between moderate redshift GAMA galaxies and high-redshift submillimeter H-ATLAS galaxies are fit to the weak lensing prediction down to the arcmin scale. The signal is interpreted under the halo model, which is modified to accommodate massive neutrinos. We discuss the impact of the choice of cosmological parametrization on the sensitivity to neutrino masses.
   }
   {The currently available  data on the magnification bias affecting submillimeter galaxies are sensitive to neutrino masses when a cosmological parametrization in terms of the primordial amplitude of the power spectrum $(A_s$) is chosen over the local root mean square of smoothed linear density perturbations $(\sigma_8$). A clear upper limit on the sum of neutrino masses can be derived if the value of $A_s$ is either fixed or assigned a narrow Gaussian prior, a behavior that is robust against changes to the chosen value. }
   {}


   \maketitle
%

Astronomy
\section{Introduction}

Observations of neutrino oscillations \cite[see][for a review of the main experimental data]{GONZALEZGARCIA16} have made it clear that the standard model of particle physics should be extended in order to accommodate massive neutrinos. However, the absolute mass scale of these particles cannot be probed through flavor oscillation experiments, because these are only sensitive to the squared mass differences of the mass eigenstates\footnote{The observed neutrino flavors $\nu_{\alpha}$ $(\alpha=e,\mu,\tau)$, i.e., the states that couple via charged currents to leptons, are unitary combinations of the so-called mass eigenstates $\nu_i$ $(i=1,2,3)$, the ones that can be formally assigned a mass, $m_{\nu_i}$, and that form the basis that diagonalizes the mass matrix in the corresponding Lagrangian.}. Complementary efforts are thus needed to obtain additional information, and examples are the kinematic analysis of the $\beta$ decay \citep{MERTENS16} and the search for the neutrinoless double-$\beta$ decay \citep{DOLINSKI19}, which probe effective masses involving the mixing angles, Majorana charge conjugation parity-violating phases, and the physical masses.

In the last few years, cosmology has been recognized as a powerful and independent tool for constraining the neutrino global mass scale. Indeed, the cosmic microwave background (CMB) anisotropy power spectra, galaxy clustering, and weak cosmological lensing are all cosmological
observables that depend (at least to first order) on the sum of neutrino masses; hence their relevance for the characterization of these particles and the minimal extensions of the $\Lambda$ cold dark matter ($\Lambda$CDM) cosmological model. In particular, CMB anisotropies reign supreme among all cosmological probes in terms of their power to constrain neutrino mass. For instance, under a $\Lambda$CDM model with massive neutrinos, the \emph{Planck} analysis of the CMB temperature anisotropy power spectrum alone yielded a 95\% upper limit of $\sum m_{\nu}<0.54$ eV, which was improved to $\sum m_{\nu}<0.24$ eV when the polarization signal was considered \citep{PLANCKVI20}. This upper bound can be further decreased to $\sum m_{\nu}<0.12$ eV when baryon acoustic oscillations are additionally taken into account \citep{PLANCKVI20}.

However,  it has been argued that these constraints are  potentially subject to the effect of the so-called lensing anomaly \citep{CALABRESE08} present in the CMB power spectra. Indeed, the analysis of \emph{Planck} data yielded a value for the amplitude of CMB lensing
that was higher than expected \citep{PLANCKVI20,DIVALENTINO20}, which could bias results on neutrino masses toward smaller values \citep{CAPOZZI21}. This motivates either the analysis of CMB measurements that do not suffer from this anomaly \citep{DIVALENTINO22} or the search for additional and independent cosmological probes that are sufficiently sensitive to neutrino masses. 

Recent results from other CMB experiments, such as the South Pole Telescope \citep{DUTCHER21} or the Atacama Cosmology Telescope \citep{AIOLA20,MADHA23}, have yielded relatively stringent upper limits on $\sum m_{\nu}$, but usually through a combination with \emph{Planck} or Wilkinson Microwave Anisotropy Probe data, with the exception of the results from the SPT \citep{balkenhol21}, where a 95\% upper limit of $m_{\nu}<0.30$ eV was found from a joint analysis with baryon acoustic oscillation data. In regards to weak lensing, experiments such as the Kilo-Degree Survey \citep{DEJONG13} or the Dark Energy Survey \citep{DES16}, which perform cosmic shear measurements along with galaxy clustering and galaxy--galaxy lensing, usually find either very loose constraints  on neutrino masses \citep{TROSTER21} or no constraints at all when the data are not combined with external probes. However, a joint analysis of the Dark Energy Survey Year 3 results with baryon acoustic oscillations, redshift-space distortions, IA supernovae, and \emph{Planck} CMB data yielded an upper limit of $\sum m_{\nu}<0.20$ eV at 95\% \citep{ABBOTT23}. 

The submillimeter galaxy magnification bias has recently been proposed as a novel approach to constrain cosmology via a weak-lensing-induced cross-correlation between a foreground galaxy sample and a background set of submillimeter galaxies \citep{BON20,GON21,BON21}. Indeed, the phenomenon of magnification bias \citep[see][and references therein]{BARTELMANN01} can boost the flux of background sources while increasing the solid angle they subtend. However, imposing a flux threshold effectively creates a mismatch between the two effects, which results in an excess of background sources around those in the foreground  with respect to the absence of lensing. Although traditionally deemed inferior to shear analyses for the probing of the galaxy-matter cross-correlation, the realization that submillimeter galaxies provide an optimal background sample for magnification bias studies \citep[as shown by the very significant early detections of this cross-correlation in][]{WANG11,GON14} has turned this observable into an emerging independent cosmological probe. Therefore, the present paper proposes the use of the submillimeter galaxy magnification bias to constrain the sum of neutrino masses. This observable has been shown to be relatively sensitive to cosmological parameters such as $\Omega_m$ and $\sigma_8$, although by itself it does not currently seem to be able to constrain others, such as the baryon density parameter, $\Omega_b$, or the spectral index of the primordial power spectrum, $n_s$. However, the realization that choosing a cosmological parametrization based on $A_s$ rather than one in terms of $\sigma_8$ can induce very different sensitivities to neutrino masses (while keeping the observable insensitive to the aforementioned unconstrained parameters) motivates us to study its potential to provide a bound on $\sum m_{\nu}$. To this end, we measured the angular cross-correlation function between a sample of high-redshift background submillimeter galaxies from H-ATLAS \citep{PILBRATT10,EALES10} and a sample of moderate-redshift foreground galaxies from GAMA II \citep{DRIVER11,BALDRY10,BALDRY14,LISKE15}. Assuming a flat $\Lambda$CDM cosmology with massive neutrinos, we adopted a modified halo model for the nonlinear galaxy-matter cross-power spectrum to derive the posterior probability distribution of the sum of neutrino masses as well as additional halo occupation distribution (HOD) and cosmological parameters.

This paper is structured as follows. Section 2 lays out the theoretical background of this work. We discuss the subtleties that arise with the introduction of massive neutrinos in the cosmological setting and describe in detail how the halo model of structure formation should be modified to account for neutrino masses. The methodology we followed is described in Section 3, where we present the galaxy samples, the procedure we used to estimate the angular cross-correlation function, and the MCMC setup we used to sample the posterior probability distribution of the parameters involved. Section 4 presents the results we obtained and in Section 5 we summarize our conclusions.

\section{Theoretical framework}

\subsection{Neutrino masses in the cosmological setting}


Accommodating neutrino masses into the cosmological framework requires an extension of the minimal six-parameter $\Lambda$CDM model. The Friedmann equation for a flat $\Lambda$CDM universe in the presence of massive neutrinos reads
\begin{equation*}
    \frac{H^2(z)}{H_0^2}=\bigg[(\Omega_{\text{cdm}}+\Omega_{\text{b}})(1+z)^3+\Omega_{\gamma}(1+z)^4+\Omega_{\Lambda}+\frac{\rho_{\nu}(z)}{\rho_{\text{crit,0}}}\bigg],
\end{equation*}
where
\begin{equation*}
    \Omega_i\equiv\frac{\rho_{i,0}}{\rho_{\text{crit,0}}}
\end{equation*} 
is the ratio of the energy density of cosmological species at present $i$ to the critical density, where $i$ runs across cold dark matter (cdm), baryons (b), photons $(\gamma)$, a cosmological constant ($\Lambda$), and neutrinos $(\nu),$ and $\rho_{\nu}(z)\equiv \sum_j \rho_{\nu_j}(z)$, where the index $j$ denotes each neutrino mass eigenstate. As neutrinos are the only known particles that undergo a nonrelativistic transition, the redshift scaling of their energy density depends on their mass. Indeed, a neutrino of mass $m_{\nu_i}$ can be shown to become nonrelativistic at a redshift \citep{LESGOURGUES13}
\begin{equation*}
    1+z_{\text{nr}_i}\approx \frac{m_{\nu_i}}{5.28\cdot10^{-4}\,\text{eV}},
\end{equation*}
after which it will contribute to the energy density of matter. Given the experimental data from flavor oscillations \citep{BILENKY16}, at least two out of the three neutrino states have been nonrelativistic for a long time, which means that the Friedmann equation can be written at our redshifts of interest $(0.2-0.8)$ as

\begin{equation*}
    \frac{H^2(z)}{H_0^2}=\bigg[(\Omega_{\text{cdm}}+\Omega_{\text{b}}+\Omega_{\nu})(1+z)^3+\Omega_{\gamma}(1+z)^4+\Omega_{\Lambda}\bigg].
\end{equation*}

Indeed, even if the third neutrino was very light and thus still relativistic at these redshifts, its contribution to $\Omega_{\gamma}$ would be negligible, and therefore the above equation is a very good approximation.

Relating the neutrino density parameter, $\Omega_{\nu}$, to the sum of their mass, $\sum m_{\nu}\equiv \sum_i m_{\nu_i}$, requires careful consideration of their thermal history (taking into account that their decoupling from the cosmic plasma is not instantaneous given the proximity in time to $e^+e^-$ annihilation); this yields \citep{LESGOURGUES13}
\begin{equation}
    \Omega_{\nu}h^2=\frac{\sum m_{\nu_i}}{93.14\text{ eV}},
\end{equation}
which is a very good approximation for the same reason as in the previous paragraph. As cosmological observables depend on the sum of neutrino masses at first order \citep{LESGOURGUES13}, the minimal extension of the $\Lambda$CDM model that includes neutrino masses is thus a mixed dark-matter model with $\sum m_{\nu}$ as an additional parameter.

The cosmological effects that arise from the introduction of neutrino masses  ---beyond the background evolution---  have been extensively studied \citep{LESGOURGES06,LESGOURGUES13,COSTANZI13,CASTORINA14,LOVERDE14,MASSARA14}. The small (but nonzero) mass of neutrinos induces very large thermal velocities, which in turn sets a free streaming scale that, in principle, prevents their clustering within dark matter halos. In reality, neutrinos from the low-velocity tail of the momentum distribution are able to cluster within the potential well of a cold dark matter halo. The strength and scale of the clustering is set by the mass of the neutrinos, the mass of the cold-dark-matter host halo, and redshift \citep{SINGH03,RINGWALD04,BRANDBYGE10,VILLANAV13}.

In cosmological terms, neutrino-free streaming leads to a slowdown and a suppression of the growth of matter perturbations on scales smaller than the free-streaming scale \citep{LESGOURGES06}. Therefore, power spectra involving the matter field are modified with respect to the absence of massive neutrinos, and the effect, which depends at linear order on the sum of neutrino masses, will transfer onto observables probing the correlation of matter with itself or with other tracers (such as galaxies, as in our case). While the effect can be quantified analytically in the linear regime, N-body simulations are usually needed to describe the matter and neutrino power spectra in the nonlinear regime \citep{VIEL10,VILLANAV13,CASTORINA15,ZENNARO19}. Another possibility, first introduced by \citet{MASSARA14}, is the analytical description (in the presence of massive neutrinos) of the power spectra in the nonlinear regime using the halo model \citep{COORAY02,ASGARI23}, which is the procedure followed throughout this paper, as detailed in the following subsection.

\subsection{The cross-correlation function and the halo model}

As explained in \cite{BON20}, \cite{GON21}, and \cite{CUE22}, the phenomenon of magnification bias probes the galaxy-mass correlation via the weak-lensing-induced angular cross-correlation function between two samples of galaxies with nonoverlapping redshift distributions. Under the Limber and flat-sky approximations, this can be expressed as \citep{COORAY02}

\begin{align}
    w_{\text{fb}}(\theta)&=2(\beta-1)\int_0^{\infty}\frac{dz}{\chi^2(z)}\frac{dN_{\text{f}}}{dz}W^{\text{lens}}(z)·\nonumber\\
    &\int_0^{\infty}dl\frac{l}{2\pi}P_{\text{g-m}}\,(l/\chi(z),z)\,J_0(l\theta)\label{crosscorr}.
\end{align}
In the above equation,
\begin{equation}
    W^{\text{lens}}(z)\equiv \frac{3}{2}\frac{1}{c^2}\bigg[\frac{H(z)}{1+z}\bigg]^2\int_z^{\infty}dz'\frac{\chi(z)\chi(z'-z)}{\chi(z')}\frac{dN_{\text{b}}}{dz'},
\end{equation}
where $dN_{\text{b}}/dz$ ($dN_{\text{f}}/dz$) is the unit-normalized background (foreground) source distribution, $\chi(z)$ is the co-moving distance at redshift $z$, $J_0$ is the zeroth-order Bessel function of the first kind, and $\beta$ is the logarithmic slope of the background source number counts.

The galaxy-matter cross power spectrum, $P_{\text{g-m}}(k,z)$, has been computed within the halo model formalism, as in previous related works. However, in contrast to these works, which assumed a minimal flat $\Lambda$CDM cosmology, modifications to the underlying framework are needed for this paper given the existence of a number of subtleties in the massive neutrino setup, as explained below.

As the matter overdensity field is a weighted average of the cold dark matter + baryon and the neutrino fields, the galaxy-matter cross-power spectrum is given by
\begin{equation}
    P_{\text{g-m}}(k,z)=(1-f_{\nu})P_{\text{g-c}}(k,z)+f_{\nu}P_{\text{g-$\nu$}}(k,z),
    \label{Pgm}
\end{equation}
where $f_{\nu}=\Omega_{\nu}/\Omega_m$ and "c" denotes the "cold" field (i.e., the weighted average of the cold dark matter and baryon fields). We now turn to the computation of each of the two terms in the above equation within the halo model, although the dominant contribution comes from the first one.

Following the discussion of \cite{MASSARA14}, both the halo mass function and the linear halo bias ---which are crucial ingredients of the halo model--- need to be modified to account for neutrino masses. Indeed, it is not clear a priori which field (i.e., the total matter field or only the cold field) should be used to define the number density of cold halos. However, the fact that only the cold field should be used is supported by the works of \cite{CASTORINA14} and \cite{VILLAESCUSA14}, which show that following this prescription implies a more universal halo mass function and a more scale-independent halo bias at large scales, respectively. Furthermore, the total-matter power spectrum was better reproduced in \cite{MASSARA14} when using ---following their nomenclature--- this "cold dark matter prescription" over that involving the full matter field. This is due to the fact that very few neutrinos are bound by cold halos, which explains why including them all via the so-called matter prescription yields different results. The cold halo mass function is given by
\begin{equation*}
    n(M_c,z)=\frac{\bar{\rho}_{\text{c}}}{M_c^2}f(\nu)\Bigg|\frac{d\log{\nu}}{d\log {M_c}}\Bigg|,
\end{equation*}
where $\bar{\rho}_{\text{c}}$ is the co-moving cold dark matter + baryon density, $f(\nu)$ defines the chosen halo mass function model, and
\begin{equation*}
    \nu(M,z)\equiv \bigg[\frac{\hat{\delta}_{\text{crit}}(z)}{\sigma_{\text{c}}(M,z)}\bigg]^2.
\end{equation*}
In the above expression, $\hat{\delta}_{\text{crit}}(z)$ is the linear critical overdensity at redshift $z$, computed via the fit of \cite{KITAYAMA96}, and
\begin{equation}
    \sigma^2_{\text{c}}(M,z)\equiv\int_0^{\infty}\frac{dk}{2\pi^2}k^2\,\tilde{W}^2(kR)\,P_{c}^{\text{lin}}(k,z)
\end{equation}
is the variance of the smoothed linear cold overdensity field, where $\hat{W}$ is the Fourier transform of the filter function (taken to be a top hat in real space) and $P^{\text{lin}}_{\text{c}}(k,z)$ is the linear cold matter power spectrum computed via the Boltzmann code CLASS \citep{BLASS11}. 

Given the above discussion, the halo model prescription for the cold matter power spectrum for the galaxy reads
\begin{equation*}
P_{\text{g-c}}(k,z)=P_{\text{g-c}}^{\text{1h}}(k,z)+P_{\text{g-c}}^{\text{2h}}(k,z)   ,
\end{equation*}
where
\begin{align*}
    P_{\text{g-c}}^{\text{1h}}(k,z)&=\int_0^{\infty} dM_c\,n(M_c,z)\frac{M_c}{\bar{\rho}_c}\frac{\langle N_{c}\rangle_{M_c}}{\bar{n}_g(z)}|u_c(k|M_c,z)|+\\
    &+\int_0^{\infty} dM_c\,n(M_c,z)\frac{M_c}{\bar{\rho}_c}\frac{\langle N_{s}\rangle_{M_c}}{\bar{n}_g(z)}|u_g(k|M_c,z)||u_c(k|M_c,z)|,
\end{align*}
and
\begin{align*}
    P_{\text{g-c}}^{\text{2h}}&(k,z)=P_c^{\text{lin}}(k,z)\Bigg[\int_0^{\infty}dM_c\,M_c\frac{n(M_c,z)}{\bar{\rho}_c}b_1(M_c,z)u_c(k|M_c,z )\Bigg]\,\nonumber\\
    &\Bigg[\int_0^{\infty}dM_c\frac{n(M_c,z)}{\bar{n}_g(z)}b_1(M_c,z)\,\Big(\langle N_c\rangle_{M_c} + \langle N_s \rangle_{M_c} \,u_g(k|M_c,z)\Big)\Bigg].
\end{align*}

In the above expressions, $b_1(M_c,z)$ is the linear halo bias (computed from the halo mass function via the peak-background split), $\bar{n}_g(z)$ is the mean galaxy number density, and 
\begin{equation*}
    \langle N \rangle_{M_c}=\langle N_{\text{cen}} \rangle_{M_c}+\langle N_{\text{sat}} \rangle_{M_c}=\bigg[1+\bigg(\frac{M_c}{M_1}\bigg)^{\alpha}\bigg]\,\Theta(M_c-M_{\text{min}})\label{HOD}
\end{equation*}
is the mean number of galaxies in a cold halo of mass $M_c$, split into the contribution of central and satellites according to the three-parameter model of \cite{ZEHAVI05}. Moreover, $u_c$ and $u_g$ are the normalized Fourier transforms of the cold dark matter halo density profile and the satellite galaxy distribution, respectively, which are assumed to follow Navarro-Frenk-White models.

Regarding the computation of the second term in \eqref{Pgm}, that is, the galaxy-neutrino cross-power spectrum, it can be further decomposed following \cite{MASSARA14} by splitting the neutrino density field into the "unclustered" ---or linear theory ($\delta_{\nu}^{\text{L}}$)--- and clustered ($\delta_{\nu}^{h}$) neutrino fields, that is, 

\begin{equation*}
   P_{\text{g-m}}(k,z)=(1-f_{\nu})P_{\text{g-c}}(k,z)+f_{\nu}\Bigg[(1-F_h)P^{\text{L}}_{\text{g}-\nu}(k,z)+F_h\,P_{\text{g}-\nu}(k,z)\Bigg], 
\end{equation*}

where $F_h$ denotes the fraction of neutrinos that do cluster within cold halos. As both $f_{\nu}$ and $F_h$ are small\footnote{The typical value for $F_h$ is $\sim 10^{-3}$ \citep{MASSARA14}. }, it is clear that the galaxy-matter power spectrum is dominated by the cross-power spectrum between galaxies and the cold field, followed by a subdominant contribution from the cross-power spectrum between galaxies and the linearly evolving neutrino field and a negligible term from the neutrino cross-power spectrum for the galaxy. Of the last two, the former can be approximated as 
\begin{equation*}
    P_{\text{g}-\nu}^{\text{L}}(k,z)\approx\sqrt{P_{gg}(k,z)P_{\nu\nu}^{
\text{L}}(k,z)},
\end{equation*}

and computed via the halo model and the Boltzmann code CAMB \citep{LEWIS11}. The latter could in principle be written within the massive neutrino halo model formalism, for which several assumptions and a neutrino density profile are needed. As this term was negligible for our purposes, its contribution was ignored.

\section{Methodology}

\subsection{Galaxy samples}

The foreground and background galaxy samples were extracted from the GAMA II \citep{DRIVER11,BALDRY10,BALDRY14,LISKE15} and H-ATLAS \citep{PILBRATT10,EALES10} surveys, respectively. Their common area covers three regions on the celestial equator at 9, 12, and 14.5 h (G09, G12, and G15) and part of the south Galactic pole, which amounts to a total of $\sim$ 207 deg$^2$. These are the same samples used in most previous related works \citep[see][and references therein for further details]{BON24}.

\begin{figure}[h]
    \centering
    \includegraphics[width=\columnwidth]{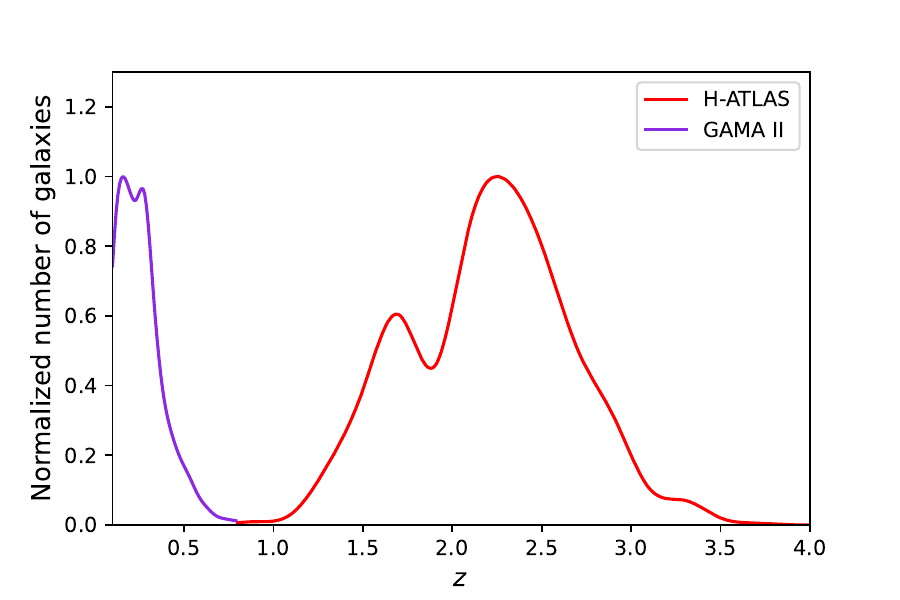}
    \caption{Normalized redshift distribution of the background (in purple) and foreground (in red) samples of galaxies.}
    \label{reddist}
\end{figure}

The foreground sample is made up of GAMA II sources with spectroscopic redshifts in the range $0.2<z<0.8$, resulting in $\sim$ 130000 galaxies with a median redshift of 0.28 and surveyed down to an r-band magnitude of 19.8. The associated redshift distribution is shown in purple in Fig. \ref{reddist}. The background sample is made up of $\sim$ 37000 H-ATLAS reliably detected sources obtained via a photometric redshift selection of $1.2<z<4.0$ to ensure that there is no overlap with the foreground galaxies. The redshift distribution of the background sample, taking random errors into account, is shown in Fig. \ref{reddist} (in red).

\subsection{Cross-correlation measurement}

\begin{figure}[h]
    \centering
    \includegraphics[width=0.9\columnwidth]{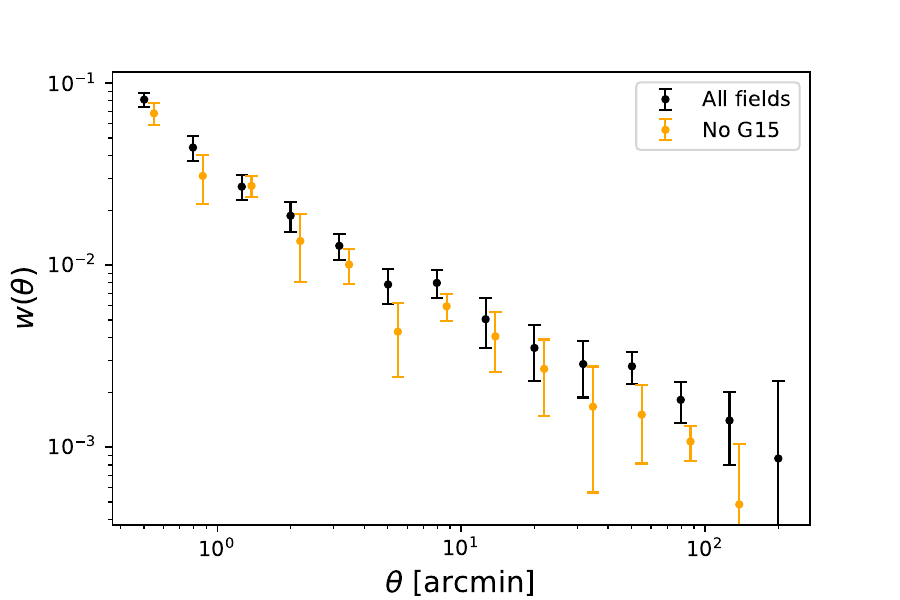}
    \caption{Angular cross-correlation measurements using all four fields (black) and excluding the G15 region (orange). }
    \label{xc_data}
\end{figure}

The measurement procedure is explained in detail in \cite{CUE24a}, but we briefly summarize the idea below. We performed a sole measurement of the angular cross-correlation function over the entire available area via the natural modification of the Landy-Szalay  estimator \citep{LANDY93}:
\begin{equation}
    \tilde{w}(\theta)=\frac{\text{D}_{\text{f}} \text{D}_{\text{b}}(\theta)-\text{D}_{\text{f}}\text{R}_{\text{b}}(\theta)-\text{D}_{\text{b}}
    \text{R}_{\text{f}}(\theta)+\text{R}_{\text{f}}^i \text{R}_{\text{b}}(\theta)}{\text{R}_{\text{f}}\text{R}_{\text{b}}(\theta)},
\end{equation}

where $\text{X}_{\text{f}} \text{Y}_{\text{b}}(\theta)$ is the number of foreground-background galaxy pairs at an angular distance $\theta$; when X$\equiv$D, the galaxies are chosen from the data, whereas X$\equiv$R implies that the sources are selected from a randomly (unclustered) generated catalog. The associated covariance matrix was computed internally through a Bootstrap resampling method with an oversampling factor of 3 \citep{NORBERG09}, that is,

\begin{equation}
    \text{Cov}(\theta_{i}, \theta_{j}) = \frac{1}{N_{b}-1}\sum_{k=1}^{N_{b}}[\hat{w_{k}}(\theta_{i})-\overline{\hat{w}}(\theta_{i}][\hat{w_{k}}(\theta_{j})-\overline{\hat{w}}(\theta_{j}],
\end{equation}
where $\hat{w}_k$ denotes the cross-correlation measurement from the k$^{th}$ Bootstrap sample, $\bar{\hat{w}}$ is the corresponding average over all Bootstrap samples, and $N_{b}=10000$. The measurements are plotted in Fig. \ref{xc_data} in black.

\subsection{Parameter estimation}

A Bayesian approach was adopted for the estimation of the probability distribution of the involved parameters. We carried out several MCMC analyses using the open source \emph{emcee} software package \citep{FOREMAN13}, which is a Python implementation of the Goodman \& Weare so-called affine invariant MCMC ensemble sampler \citep{GOODMAN10}. The log-likelihood functions was assumed to be a multivariate Gaussian distribution, that is,

\begin{align}   \log{\mathcal{L}\,(\theta_1,\ldots,\theta_m)}=-\frac{1}{2}&\bigg[m\log{(2\pi)}+\log{|C|}+\overrightarrow{\varepsilon}^{\text{T}}C^{-1}\,\overrightarrow{\varepsilon}\bigg],
\end{align}
where 
\begin{equation*}  \overrightarrow{\epsilon}\equiv [\varepsilon(\theta_1),\ldots,\varepsilon(\theta_m)]
\end{equation*}

and $\varepsilon(\theta_i)=w_{\text{fb}}(\theta_i)-\hat{w}_{\text{cross}}(\theta_i)$.

\begin{table}[h]

\centering
\caption{Parameter priors used in this work.}

\begin{tabular}{c | c}
\hline
Parameter & Prior  \\
\hline
$\alpha$ & $\mathcal{U}[0.0, 1.5]$ \\
$\log M_{1}$ & $\mathcal{U}[10.0, 16.0]$\\
$\log M_{min}$ & $\mathcal{U}[10.0, 16.0]$ \\
$\Omega_{M}$ & $\mathcal{U}[0.1, 1.0]$ \\
$h$ & $\mathcal{U}[0.5, 0.9]$\\
$\sum m_{\nu}$ & $\mathcal{U}[0.0, 1.5]$\\
$\beta$ & $\mathcal{N}[2.90, 0.04]$\\
\hline
\label{priors}
\end{tabular}
\tablefoot{$\mathcal{U}[a,b]$ denotes a uniform distribution with range $[a,b]$ and $\mathcal{N}[\mu,\sigma]$ denotes a Gaussian with mean $\mu$ and standard deviation $\sigma$.}
\end{table}

\begin{figure*}[t]
    \centering
    \includegraphics[width=0.95\columnwidth]{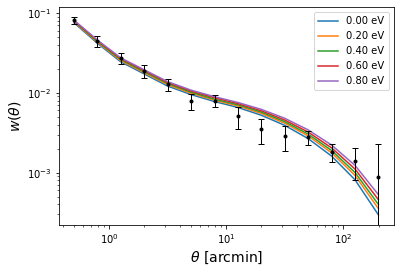}
    \includegraphics[width=0.95\columnwidth]{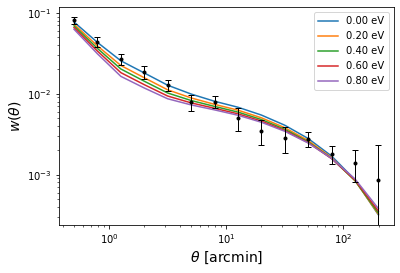}
    
    \caption{Sensitivity of the angular cross-correlation function with respect to $\sum m_{\nu}$ for fixed values of $\sigma_8=0.79$ (left panel) and $\log{(10^{10}\,A_s)}=3.35$ (right panel).}
    \label{xc_sensitivity_to_mnu_fixed_sgm8_and_As}
\end{figure*}

Several different cases were studied in this work according to the parameterization of the model (see the following section), but the estimation procedure involved the same parameters in almost all scenarios, namely the HOD ($\alpha$, $M_{\text{min}}$ and $M_1$) and a massive neutrino $\Lambda$CDM cosmology with a fixed power spectrum normalization ($\Omega_m$, $h$ and $\sum m_{\nu}$), along with the number count logarithmic slope, $\beta$. The prior distributions used for all cases are summarized in Table \ref{priors}. They are all uniform with the exception of the logarithmic slope of the background number counts, which was chosen to follow a Gaussian distribution around 2.9 following the analysis of \cite{CUE24a}.


\section{Results}

\subsection{Sensitivity to neutrino masses}

It is well known that the minimal $\Lambda$CDM cosmological model can be described through the six-parameter family $(\Omega_m,\Omega_b,n_s,\sigma_8,h,\tau)$, where $\sigma_8$ fixes the normalization of the $z=0$ linear matter power spectrum. However, this model can also be  parameterized by means of the equivalent set $(\Omega_m,\Omega_b,n_s,A_s,h,\tau)$, where $A_s$ is the amplitude of the primordial spectrum of curvature perturbations at a pivot scale $k_0$, that is\footnote{The pivot scale is taken to be $k_0\equiv0.05$ Mpc$^{-1}$ throughout the present paper.}
\begin{equation*}
    \frac{k^3}{2\pi^2}P_{\zeta}(k)=A_s\,\Bigg(\frac{k}{k_0}\Bigg)^{n_s-1}.
\end{equation*}

These two normalization parameters $(\sigma_8$ and $A_s$) can be linked given the relation between the linear matter spectrum and the primordial spectrum of curvature perturbations \citep{DODELSON20}:
\begin{equation*}
    P_{\text{mm}}^{\text{lin}}(k,z)=\frac{8\pi^2}{25} \frac{A_s\,k_0}{\Omega_m^2}D^2(z)T^2(k,z)\,\Bigg(\frac{k}{k_0}\Bigg)^{n_s}\Bigg(\frac{c}{H_0}\Bigg)^4.
\end{equation*}

In the above equation, the transfer function $T(k,z)$ satisfies $T(k\to0,z)\to 1$ and is redshift-dependent in the massive-neutrino setup\footnote{Equivalently, one could also say the linear growth factor becomes scale dependent in the massive-neutrino setup.}. In turn, the linear growth factor is given by

\begin{equation}
    D(z)=\frac{5}{2}\Omega_m \frac{H(z)}{H_0}\int_z^{\infty}\frac{1+z'}{(H(z')/H_0)^3}dz'.
\end{equation}

The addition of massive neutrinos does not modify the above equations, but affects the computation of both the growth factor and the transfer function. This entire discussion is related to the fact that, interestingly, the choice of parameter family has a large impact on the constraining power of our observable regarding neutrino masses. Indeed, the left panel of Figure \ref{xc_sensitivity_to_mnu_fixed_sgm8_and_As} shows the sensitivity of the theoretical cross-correlation function to changes in $\sum m_{\nu_i}$ of up to 0.80 eV in the $\sigma_8$ parameterization. It is clear that the variation is small and only noticeable at the largest angular scales, where the error bars are still sizeable in comparison. Therefore, an attempt to constrain neutrino masses, even by fixing the $\sigma_8$ parameter, is almost certain to produce nonconstraining results, as we show below.

\begin{table*}[t]
\centering
\caption{MCMC results from each considered case.}
\label{Table_results_all_cases}

\begin{tabular}{c   c  c  c  c  c  c  c}
\hline
\hline
& $\sigma_8$ CUE24a & $A_s$ CUE24a & $A_s$ \emph{Planck} & $A_s$ \emph{Planck} ($\sigma=0.1)$ & $A_s$ CUE24a (no G15) \\ 
\hline \vspace{-0.25cm} \\

$\alpha$ & $0.69^{+0.30}_{-0.46}\,(0.60)$ & $>0.79\,(-)$ & $>1.17 \,(-)$ & $>1.09\,(-)$ & $0.89^{+0.18}_{-0.26}\,(0.81)$\vspace{0.07cm}\\

$\log M_{\text{min}}$ & $11.64^{+0.17}_{-0.12}\,(11.68)$ & $11.47^{+0.14}_{-0.11}\,(11.50)$ & $11.49^{+0.14}_{-0.16}\,(11.50)$& $11.53^{+0.15}_{-0.14}\,(11.54)$& $11.46^{+0.27}_{-0.16}\,(11.53)$\vspace{0.07cm}\\

$\log M_1$ & $13.39^{+0.71}_{-0.92}\,(13.34)$ & $12.89^{+0.56}_{-0.41}\,(13.01)$ & $12.52^{+0.36}_{-0.25}\,(12.61)$ & $12.59^{+0.50}_{-0.34}\,(12.71)$ & $12.43^{+0.70}_{-0.46}\,(12.68)$\vspace{0.07cm}\\

$\Omega_m$ & $0.28^{+0.04}_{-0.08}\,(0.25)$ & $0.18^{+0.02}_{-0.02}\,(0.18)$ & $0.16^{+0.01}_{-0.02}\,(0.16)$ & $0.17^{+0.01}_{-0.03}\,(0.16)$ &$0.34^{+0.02}_{-0.03}\,(0.33)$\vspace{0.07cm}\\

$h$ & $<0.70\,(-)$ & $>0.77\,(-)$ & $>0.82\,(-)$ & $>0.83\,(-)$& $0.72^{+0.07}_{-0.06}\,(0.73)$\vspace{0.07cm}\\

$\sum m_{\nu}$ (eV) &$-\,(-)$& $<0.22\,(-)$ & $<0.36\,(-)$ & $<0.46\,(-)$ & $-\,(-)$\vspace{0.07cm}\\

$\beta$ & $2.91^{+0.04}_{-0.04}\,(2.91)$ & $2.91^{+0.03}_{-0.02}\,(2.91)$ & $2.92^{+0.04}_{-0.04}\,(2.92)$ & $2.92^{+0.04}_{-0.04}\,(2.92)$ &$2.90^{+0.04}_{-0.03}\,(2.90)$\vspace{0.07cm}\\

\hline
\hline
\end{tabular}
\tablefoot{The mean and 68\% credible intervals for each parameter are stated, along with the marginalized maximum a posteriori value in parenthesis. Dashes indicate unconstrained distributions or unmeaningful maximum a posteriori values. Halo masses are expressed in $M_{\odot}/h. $}
\end{table*}

The situation in the parameterization that includes $A_s$ is completely different. The right panel of Figure \ref{xc_sensitivity_to_mnu_fixed_sgm8_and_As} depicts the sensitivity of the cross-correlation function to the same variations as the previous case, but this time in the $A_s$ parameterization. The response of the observable is now substantial and predominantly takes place at intermediate angular scales, where the error bars are smaller. Although this does not necessarily mean that a full MCMC analysis including $A_s$ instead of $\sigma_8$ will compare favorably with the previous case, it does hint at the possibility of restricting neutrino masses when the value of $A_s$ is fixed. This finding is the main motivation behind the present paper.

\subsection{Cosmological analysis: $A_s$ vs $\sigma_8$}

The first case we studied involved fixing the $\sigma_8$ parameter to 0.79, the best fit from the massless neutrino case in \cite{CUE24a}, whereas the rest of parameters ($\alpha,M_{\text{min}},M_1,\Omega_m,h,\sum m_{\nu},\beta)$ were allowed to vary. The corresponding statistical results are summarized in the first column of Table \ref{Table_results_all_cases} and Fig. \ref{cornerplot_As_fixed_vs_sgm8_fixed} (in red). Concerning the HOD, we obtained results in agreement with typical values from the literature \citep{ZEHAVI05,ABBAS10,ZEHAVI11} and with the findings of the massless neutrino case from \cite{CUE24a}; we derived mean values of $\log M_{\text{min}}=11.64^{+0.17}_{-0.12}$ and $\log M_1=13.39^{+0.71}_{-0.92}$ for the halo masses and $\alpha=0.69^{+0.30}_{-0.46}$ for the satellite logarithmic slope.

\begin{figure}[h]
    \centering
    \includegraphics[width=0.9\columnwidth]{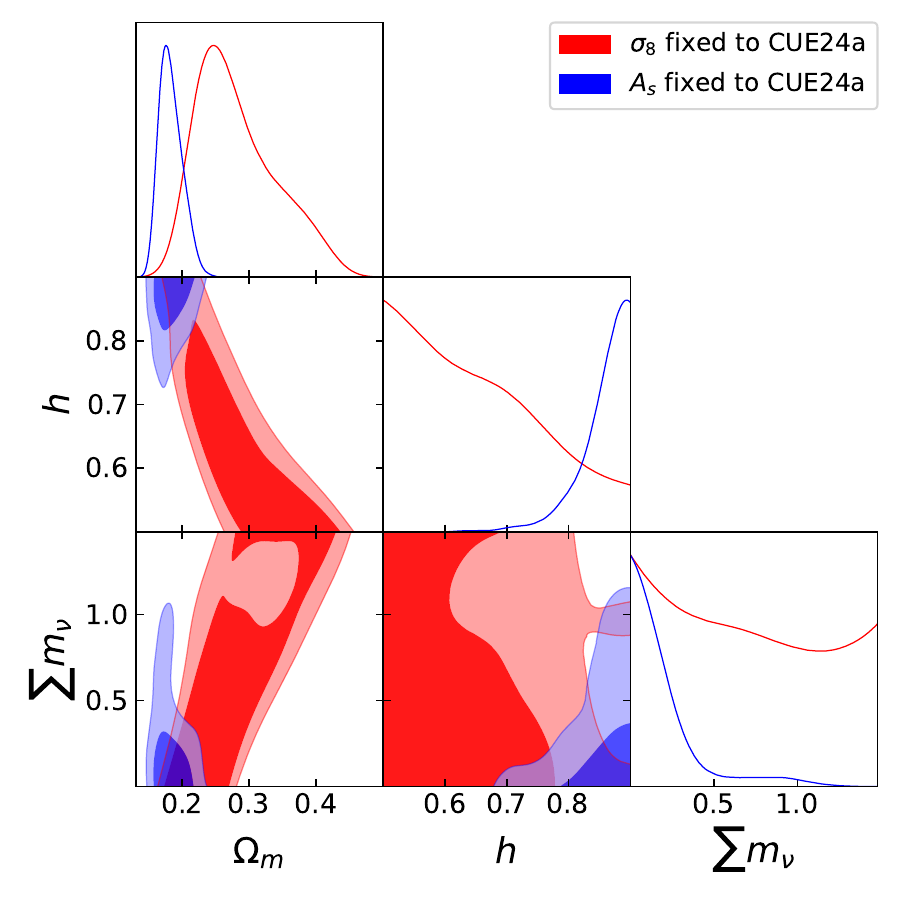}
    \caption{Marginalized posteriors and probability contours of the cosmological parameters for fixed $\sigma_8$ (in red) and $A_s$ (in blue) values according to the best fit from the massless neutrino case of CUE24a.}
    \label{corner_cosmo_fixed_sgm8_vs_fixed_As}
\end{figure}

Regarding cosmology, we obtained a mean value of $\Omega_m=0.28^{+0.04}_{-0.08}$ for the matter density parameter, higher than the previous results by \cite{CUE24a} and \cite{BON24}, where neutrinos were considered massless. We believe this displacement toward larger values is caused by the additional large-scale freedom provided by neutrino masses, since both parameters affect mainly the high-$\theta$ regime. Indeed, as seen on the $\Omega_m-\sum m_{\nu}$ plane in Fig. \ref{corner_cosmo_fixed_sgm8_vs_fixed_As} (in red), neutrino masses close to zero imply values of the matter density parameter aligned with $\sim 0.2$, as found by \cite{CUE24a} and \cite{BON24}. Only an upper limit can be found for the Hubble constant, with $h<0.70$ at 68\% and, as expected from the analysis in the previous subsection, the neutrino mass sum cannot be constrained in this scenario.

We next moved to the parametrization of the model where $\sigma_8$ is replaced by $A_s$, which, as in the previous case, is fixed to the best-fit value from the massless neutrino case, which is $2.86\times 10^{-9}$. The statistical results are summarized in the second column of Table \ref{Table_results_all_cases} and Fig. \ref{corner_cosmo_fixed_sgm8_vs_fixed_As} (in blue). The HOD masses are constrained, with mean values of $\log M_{\text{min}}=11.47^{+0.14}_{-0.11}$ and $\log M_1=12.89^{+0.56}_{-0.41}$, which are in agreement with the previous case within the error bars. The logarithmic slope of the number of satellites, $\alpha$, is however not as well constrained due to the small-scale influence of neutrino masses, but we can derive a lower limit of $\alpha>0.79$ at 68\%. It should be noted that, although the fixed parameter in each of the two scenarios has been fixed according to the same best-fit cosmology, $A_s$ and $\sigma_8$ are not in a one-to-one relation, as one is a function of all cosmological parameters and not only of the other. Therefore, the quantitative results from the two cases cannot be compared directly, because they are not equivalent given the dependence on the exact value of the normalization parameter. 

As far as cosmology is concerned, as seen in Fig. \ref{corner_cosmo_fixed_sgm8_vs_fixed_As} (in blue), we obtained a well-constrained posterior distribution for the matter density parameter, with a mean value of $\Omega_m=0.18^{+0.02}_{-0.02}$ and a lower limit of $h>0.77$ at 95\% for the Hubble constant. These findings, which are aligned with the results from \cite{CUE24a} and \cite{BON24} when all fields are considered, are clearly in tension with the results of other cosmological probes \citep{ABBOTT18,PLANCKVI20,ABBOTT20} and are further discussed in the following subsection. 

Regardless, and contrary to the prior scenario, a clear upper limit is now found for the sum of neutrino masses, with $\sum m_{\nu}<0.22 $ (0.78) eV at 68\% (95\%) credibility. This result demonstrates the main finding of this paper: by fixing the matter power spectrum normalization, current submillimeter galaxy magnification bias measurements are sensitive to neutrino masses, meaning that a cosmological paramaterization based on $A_s$ as opposed to $\sigma_8$ must be adopted. 

\begin{figure}[h]
    \centering
    \includegraphics[width=0.9\columnwidth]{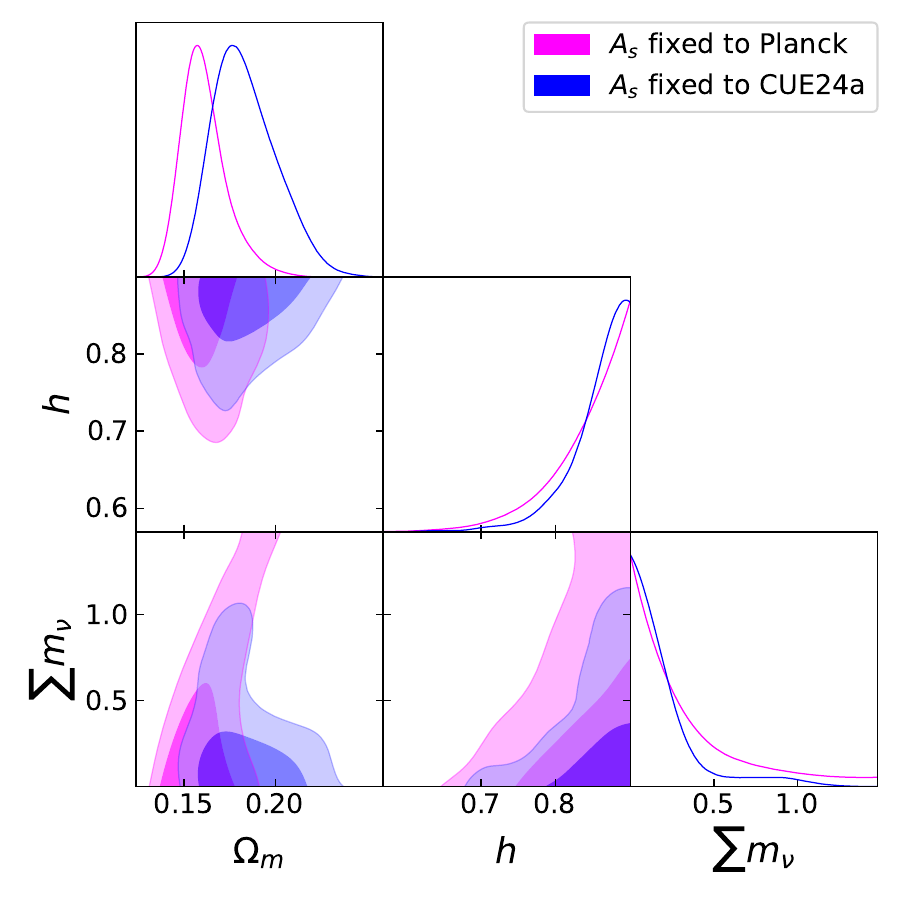}
    \caption{Marginalized posteriors and probability contours of the cosmological parameters for fixed $A_s$ values according to the best fit from CUE24a (in blue) and \emph{Planck} (in magenta).}
    \label{corner_cosmo_fixed_sAs_CUE24a_vs_Planck}
\end{figure}

\subsection{Further discussion}

In light of the results of the previous subsection, a number of aspects need to be explored regarding their validity. First, one may wonder how the above findings behave with respect to the prior distribution of $A_s$, as they have been derived assuming the best-fit value to the massless neutrino case. To investigate this matter further, we started by considering a fixed value of $A_s$ according to \emph{Planck}'s best fit (TT, EE, TE, lowE + lensing), that is, $\log(10^{10}A_s)=3.044$ \citep{PLANCKVI20}. 

The statistical results are gathered in the third column of Table \ref{Table_results_all_cases} and Fig. \ref{cornerplot_As_fixed_CUE24a_vs_Planck}. Although there are slight variations in the overall HOD, the posterior distributions of $M_{\text{min}}$ and $h$ remain unchanged. Furthermore, as seen in Fig. \ref{corner_cosmo_fixed_sAs_CUE24a_vs_Planck}, the sensitivity of the system with respect to neutrino masses stays qualitatively unaltered; indeed, a clear upper limit of $\sum m_{\nu}<0.36\, (1.08)$ eV is still found at 68\% (95\%). The matter density parameter, however, is displaced toward slightly lower values due to the degeneracy with $A_s$ (larger values of one parameter can be balanced by smaller values of the other).

\begin{figure}[h]
    \centering
    \includegraphics[width=0.9\columnwidth]{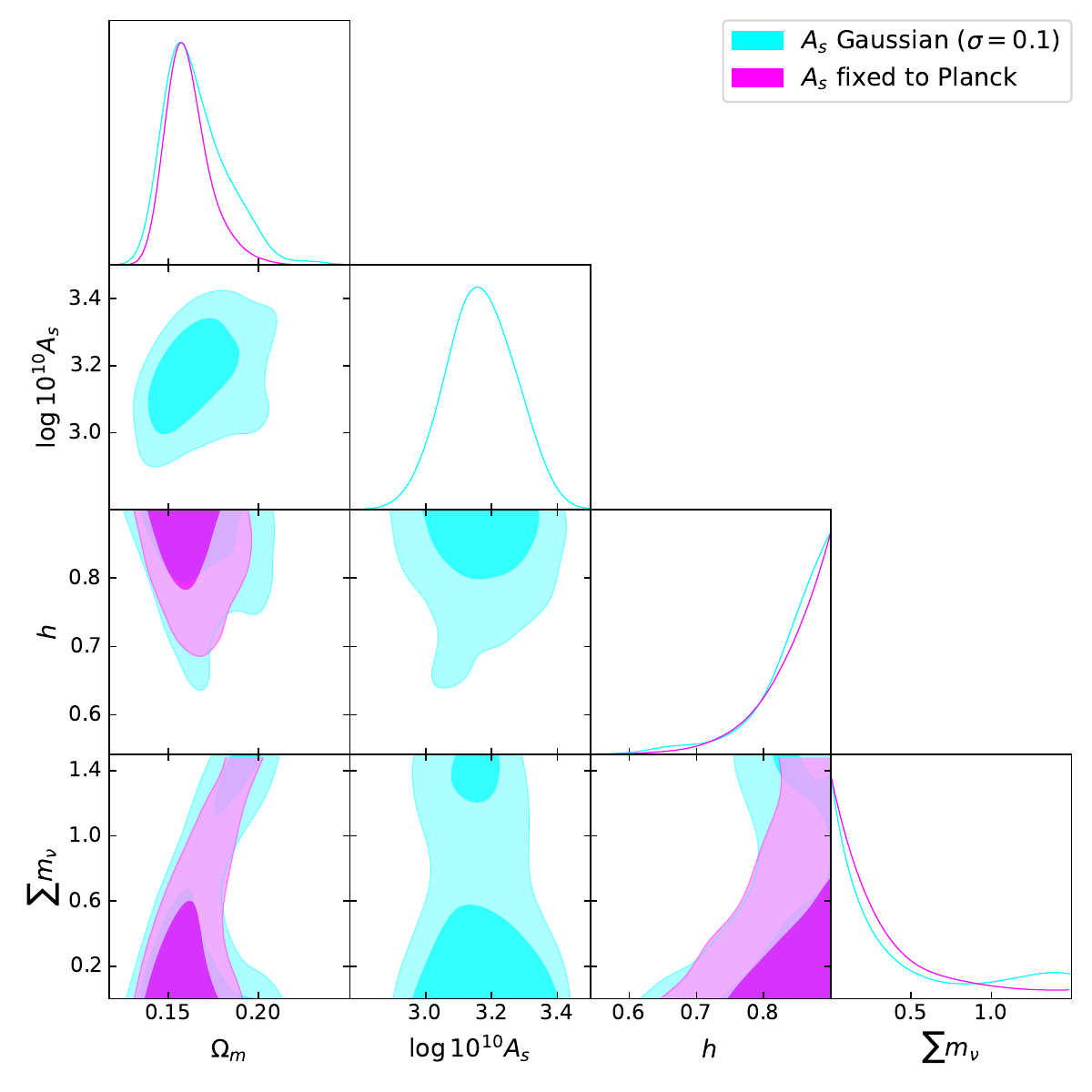}
    \caption{Marginalized posteriors and probability contours of the cosmological parameters for a fixed Planck $A_s$ value (in magenta) and for a Gaussian prior around that value (in cyan).}
    \label{corner_cosmo_fixed_vs_Gaussian_As}
\end{figure}

Second, and given the current size of the error bars, the question arises as to how wide the prior distribution of $A_s$ can be whilst maintaining the sensitivity of the model to neutrino masses. To investigate this question, we considered a Gaussian prior for $A_s$ around the above \emph{Planck} value with a standard deviation of 0.1. The results are summarized in the fourth column of Table \ref{Table_results_all_cases} and Fig. \ref{corner_cosmo_fixed_vs_Gaussian_As} in cyan. Our findings are both qualitatively and quantitatively similar to the previous case, with only slightly larger uncertainties arising from the additional degree of freedom. The posterior distribution of $A_s$ is displaced from its prior to a mean value of $\log{(10^{10}A_s)}=3.17^{+0.11}_{-0.10}$. This behavior seems to impact the sensitivity to neutrino masses, as seen more clearly in Fig. \ref{corner_cosmo_fixed_vs_Gaussian_As}; indeed, although the highest posterior density interval at 68\% is $\sum m_{\nu}<0.46$ eV, a secondary high-density region is found in the posterior at very high $(\sim 1.3$ eV) neutrino masses. However, this still indicates that neutrino masses are not prior-dominated even in the case of a relatively narrow Gaussian prior for $A_s$ and therefore that sensitivity is still manifest.

All analyses performed up to this point under the $A_s$ parametrization share a common feature: the posterior distributions of $\Omega_m$ (and, to a certain extent, that of $h$) are in tension with common values from the literature. The above tests, however, suggest that the exact prior value of $A_s$ should not be able to account for the entirety of this behavior. Indeed, we believe that the discrepant cosmological results stem from the data themselves.

As already shown in \cite{CUE24a} and \cite{BON24}, there appears to be an excess of a cross-correlation signal -especially at the largest scales - due the G15 equatorial region. As the galaxy selection criteria are uniform across all four fields (as are the redshift distributions), we hold the view that this stems from the phenomenon of sampling variance and plan to study it further with a larger sample in a follow-up work. In any case, a preliminary analysis performed in \cite{CUE24a} showed that removing the G15 region induced non-negligible changes in cosmological constraints; in particular, larger values of $\Omega_m$ were attained, which are in keeping with standard results from external probes.

\begin{figure}[h]
    \centering
    \includegraphics[width=0.9\columnwidth]{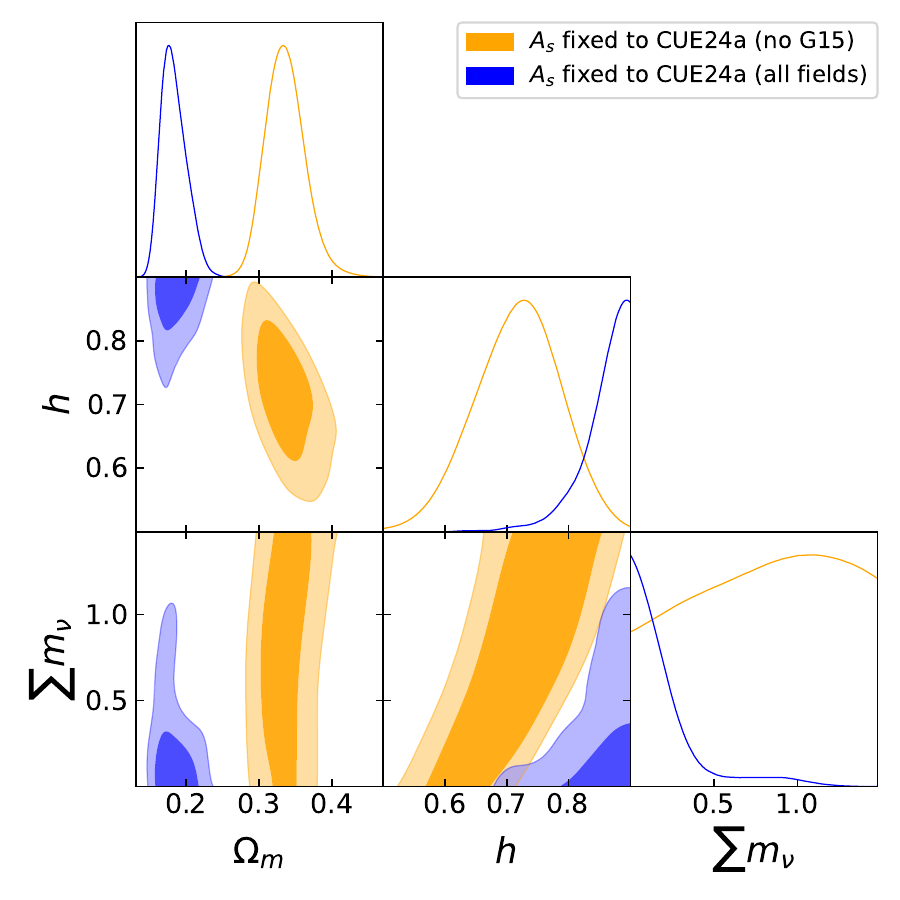}
    \caption{Marginalized posteriors and probability contours of the cosmological parameters for a fixed $A_s$ value according to the best fit from CUE24a. The results using all four fields are shown in blue, while the case where the G15 region was excluded is depicted in orange.}
    \label{corner_cosmo_fixed_As_all_vs_noG15}
\end{figure}

Therefore, we decided to assess the influence of leaving out the G15 region in our analysis with neutrino masses. To do so, we fixed $A_s$ once again to the best fit from the massless neutrino case of \cite{CUE24a}. The cross-correlation measurements for this case are shown in Fig. \ref{xc_data} in orange, where the excess of cross-correlation at intermediate and large scales is more visible when compared to all fields. 

The results are summarized in the fifth column of Table \ref{Table_results_all_cases} and Fig. \ref{cornerplot_As_fixed_all_vs_noG15}. All three HOD parameters are reasonably well constrained and their distributions are only slightly modified with respect to the four-field case. However, as shown by Fig. \ref{corner_cosmo_fixed_As_all_vs_noG15}, the posterior distribution of $\Omega_m$ is strongly displaced toward larger values, following the behavior of \cite{CUE24a} and \cite{BON24}, with a mean of $\Omega_m=0.34^{+0.02}_{-0.03}$. Moreover, the Hubble constant is constrained on its two sides, with a clear mean value of $h=0.72^{+0.07}_{-0.06}$. However, the sensitivity to neutrino masses is lost, and no constraints can be given in this scenario given the sizeable error bars arising from the exclusion of the G15 region, as confirmed by a simple parameter sensitivity analysis. As expected, the qualitative behavior of this last test remains for a fixed \emph{Planck} value of $A_s$.

The results without the G15 field are thus in tension with the findings
using all four independent regions, and this is even more noticeable than in the massless neutrino scenario. This is still a matter of ongoing investigation and will be addressed further in future work. Nonetheless, as far as neutrino masses are concerned, we find that sacrificing one field has a devastating effect on the sensitivity of cosmic magnification, although it produces cosmological results that are well in agreement with external probes. Therefore, measurement error bars must be at least the size of the four-field case, especially at the smallest scales.

Regarding the validity of the halo model, it is well known \citep{ASGARI23} that it fails to faithfully reproduce the transition between the one- and two-halo regimes, which can be alleviated by the introduction of nonlinear halo bias, as calibrated by numerical simulations \citep{NISHIMICHI19,MEAD21}.  Moreover, baryonic feedback can induce deviations to the matter and galaxy-matter power spectra with respect to the minimal halo model. Its impact on small scales can be addressed (at least partially) by the assignment of a  variable halo concentration amplitude or a separate concentration amplitude for the satellite galaxy distribution \citep{AMON23,DVORNIK23}. We will study the introduction of these effects in future work.

\section{Summary and conclusions}

This paper addresses the sensitivity of the submillimeter galaxy magnification bias to the sum of neutrino masses. We measured the angular cross-correlation function induced by the weak-lensing magnification bias between a sample of high-redshift submillimeter galaxies from H-ATLAS and a sample of moderate-redshift galaxies from GAMA. We describe the nonlinear behavior of the  cross-power spectrum of the galaxy matter via the halo model, which allows us to go down to approximately arcminute scales and describe the galaxy--halo connection via a simple three-parameter HOD model. We assume a $\Lambda$CDM cosmological model with neutrino masses, for which we implement a modified version of the halo model that entails considering only the cold dark matter + baryon field to describe the nonlinear clustering of matter.

The main goal of this study is to investigate the potential of cosmic magnification with background submillimeter galaxies as a cosmological probe of neutrino masses. We find that a cosmological paramaterization in terms of the 
amplitude of the primordial power spectrum ($A_s$) is required, rather than that involving the root mean square of the $z=0$ filtered overdensity field ($\sigma_8$). The reason is related to the different sensitivity of the angular cross-correlation to the sum of neutrino masses under each paramaterization. Indeed, the variation of the signal is concentrated on the small and intermediate scales, where the error bars are relatively small, for a fixed value of $A_s$. This is in contrast with the case of a fixed value of $\sigma_8$, where the response is located on large angular scales, rendering the current measurements insensitive to neutrino masses.

For a value of $A_s$ fixed to the best fit of the massless neutrino case of \cite{CUE24a}, we obtained clear upper bounds for the sum of neutrino masses: $\sum m_{\nu}<0.22$ eV at 68\% and $<0.78$ at 95\%. Although the exact numerical values depend on the chosen value of $A_s$, we find qualitatively stable results when it is varied.
Indeed, for a fixed \emph{Planck} value of $A_s$, we obtain $\sum m_{\nu}<0.36\,(1.08)$ eV at 68\% (95\%). Moreover, assuming a Gaussian prior for $A_s$ of around this \emph{Planck} value, we find $\sum m_{\nu}<0.46$ eV at 68\%, with a small high-density region at large neutrino masses due to the additional freedom in $A_s$. This demonstrates that, by fixing only the normalization of the primordial power spectrum (or by adopting a relatively narrow Gaussian prior distribution for it), cosmic magnification on submillimeter galaxies can be sensitive to the sum of neutrino masses.

However, we find the posterior distributions obtained for the matter density parameter to be in tension with values derived from external probes. This issue was pointed out by \cite{CUE24a} and \cite{BON24} and was deemed to be the result of an excess of cross-correlation found within the G15 region. Although the exclusion of this field solves the problem, the larger error bars due to the reduction in galaxy sample size equate to insufficient sensitivity to constrain neutrino masses. Therefore, future work will be directed toward a better understanding of this issue, enabling us to obtain both unbiased and constraining results.

\begin{acknowledgements}
LB, JGN, JMC and DC acknowledge the PID2021-125630NB-I00 project funded by MCIN/AEI/10.13039/501100011033/FEDER, UE. LB also acknowledges the CNS2022-135748 project funded by MCIN/AEI/10.13039/501100011033 and by the EU “NextGenerationEU/PRTR”. JMC also acknowledges financial support from the SV-PA-21-AYUD/2021/51301 project. MV is partly supported by the INFN IS INDARK grant. 
\end{acknowledgements}

\bibliographystyle{aa} 
\bibliography{main} 

\appendix

\onecolumn

\section{Additional plots}

\begin{figure}[h]
    \centering
    \includegraphics[width=0.9\columnwidth]{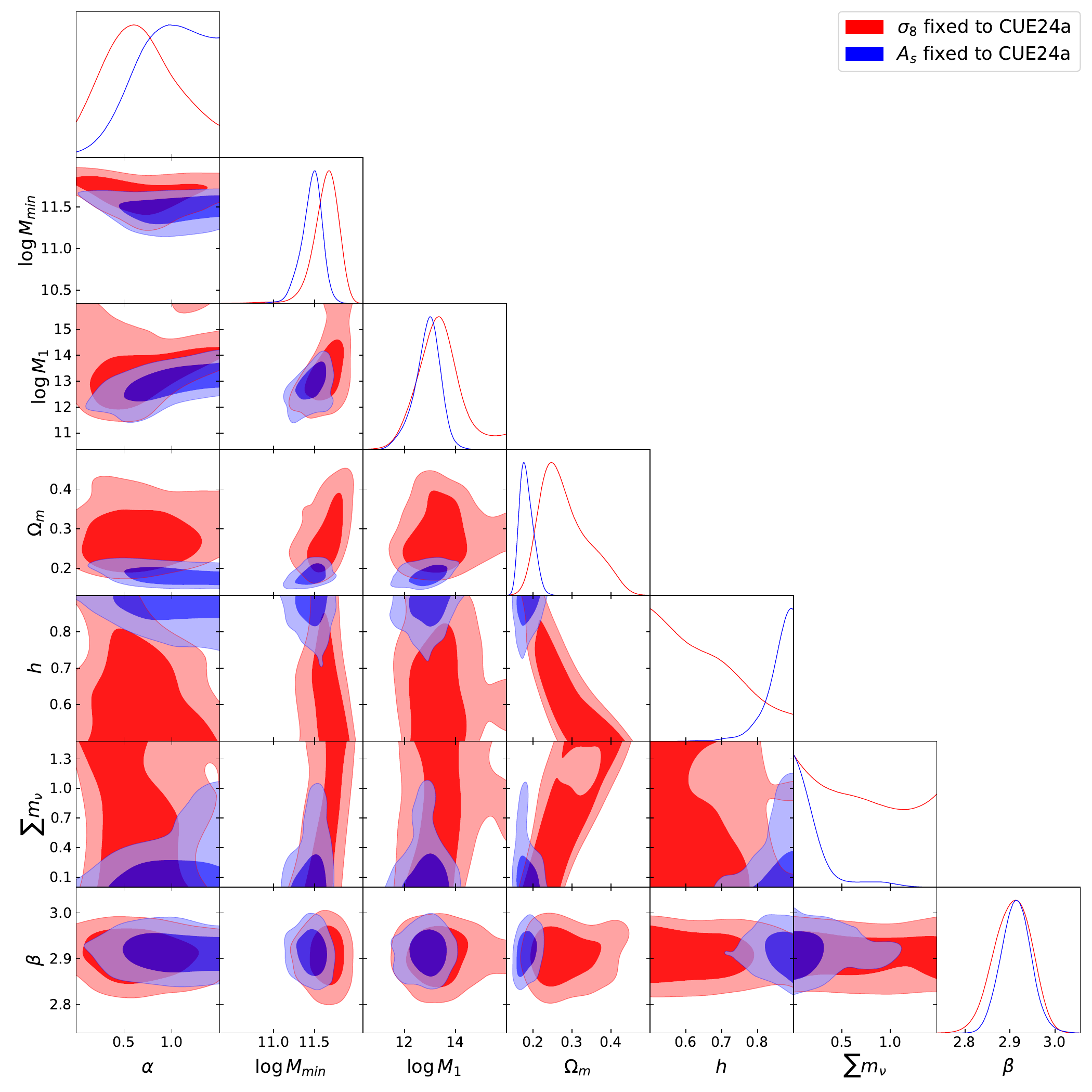}
    \caption{Marginalized posterior distributions and probability contours for fixed $\sigma_8$ (in red) and $A_s$ (in blue) values according to the best fit from the massless neutrino case of CUE24a.}
    \label{cornerplot_As_fixed_vs_sgm8_fixed}
\end{figure}

\begin{figure}[h]
    \centering
    \includegraphics[width=0.9\columnwidth]{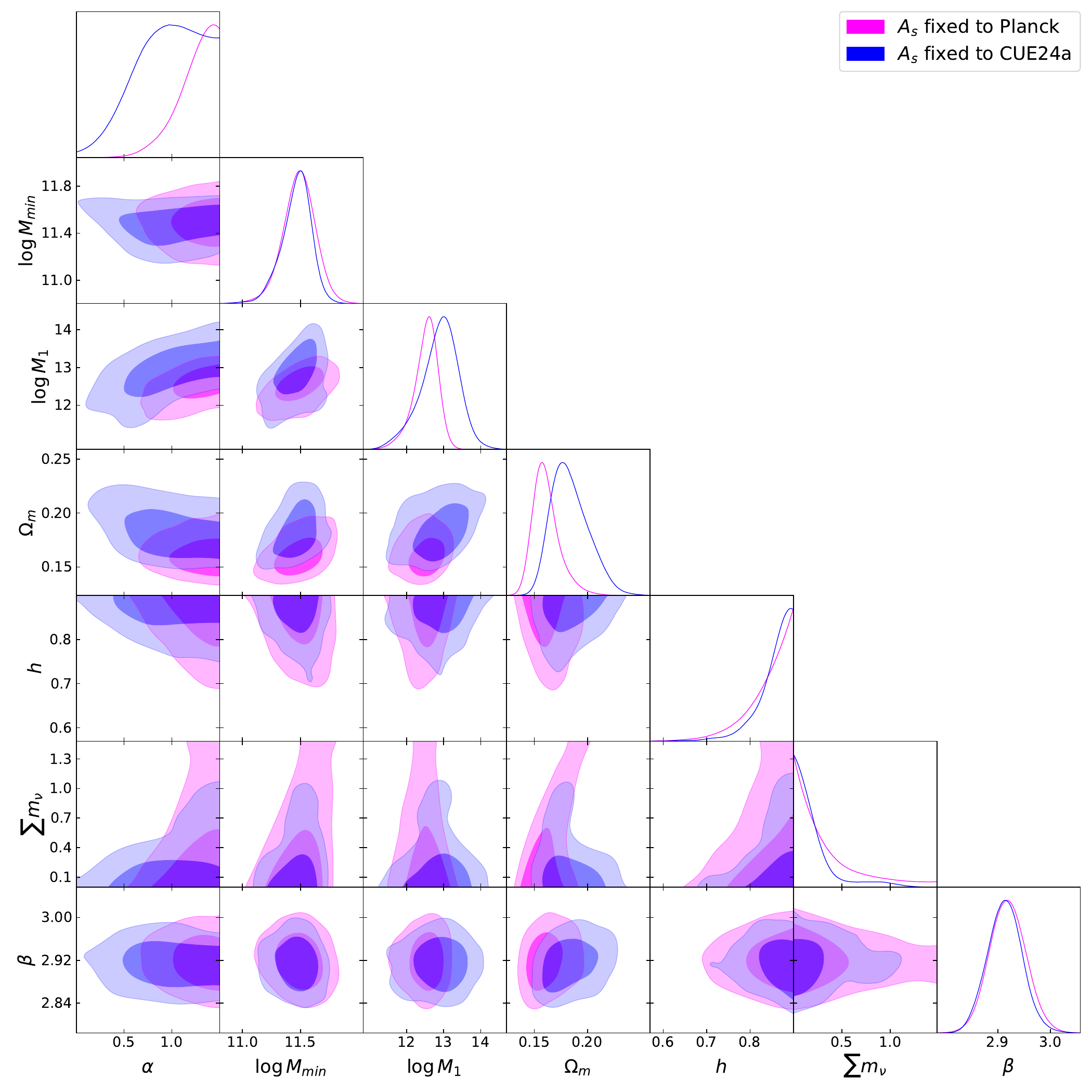}
    \caption{Marginalized posterior distributions and probability contours for a fixed $A_s$ value according to the best fit from CUE24a (in blue) and to \emph{Planck} (in magenta).}
    \label{cornerplot_As_fixed_CUE24a_vs_Planck}
\end{figure}

\begin{figure}[h]
    \centering
    \includegraphics[width=0.9\columnwidth]{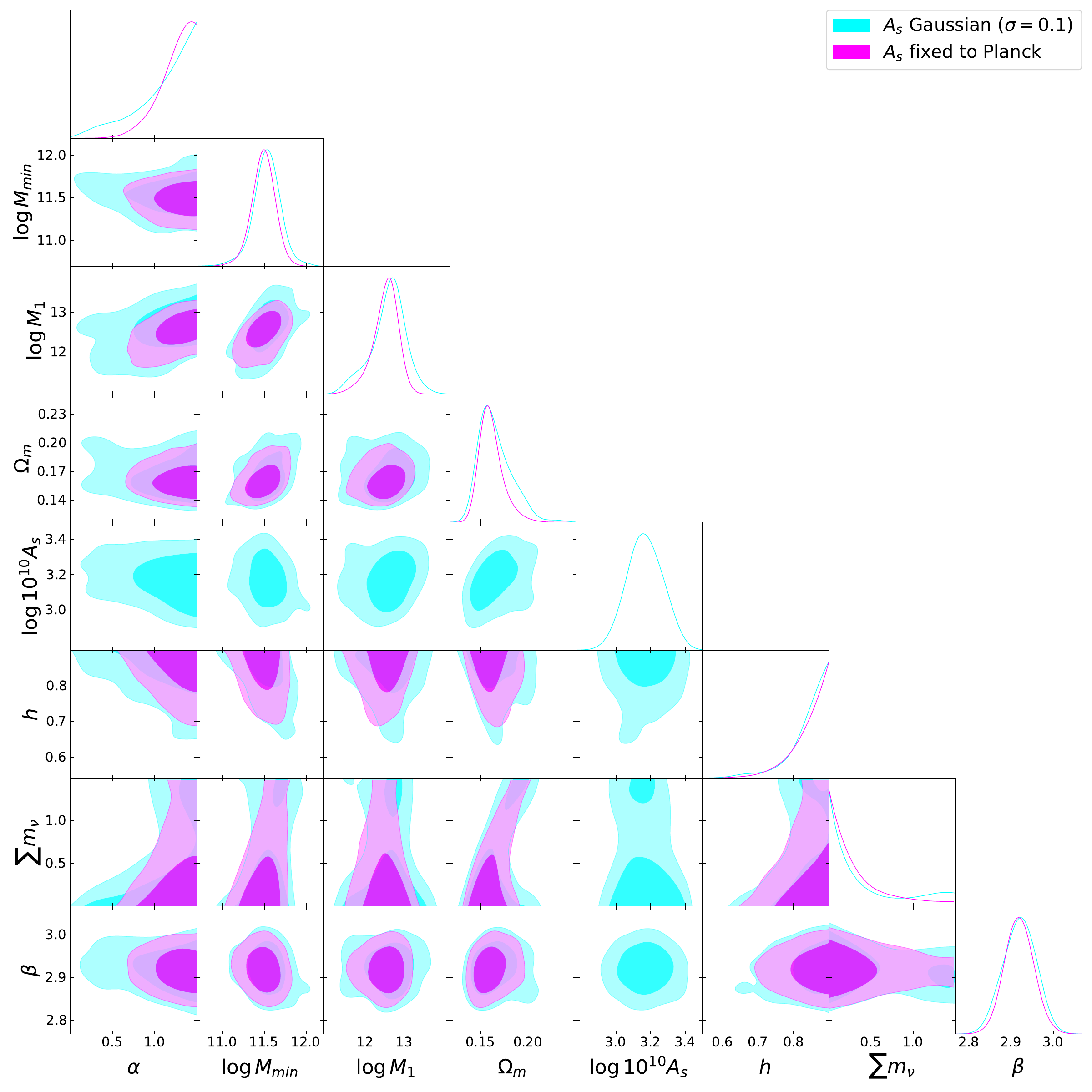}
    \caption{Marginalized posterior distributions and probability contours for a fixed Planck $A_s$ value (in magenta) and for a Gaussian prior around that value (in cyan).}
    \label{cornerplot_As_fixed_vs_Gaussian}
\end{figure}

\begin{figure}[h]
    \centering
    \includegraphics[width=0.9\columnwidth]{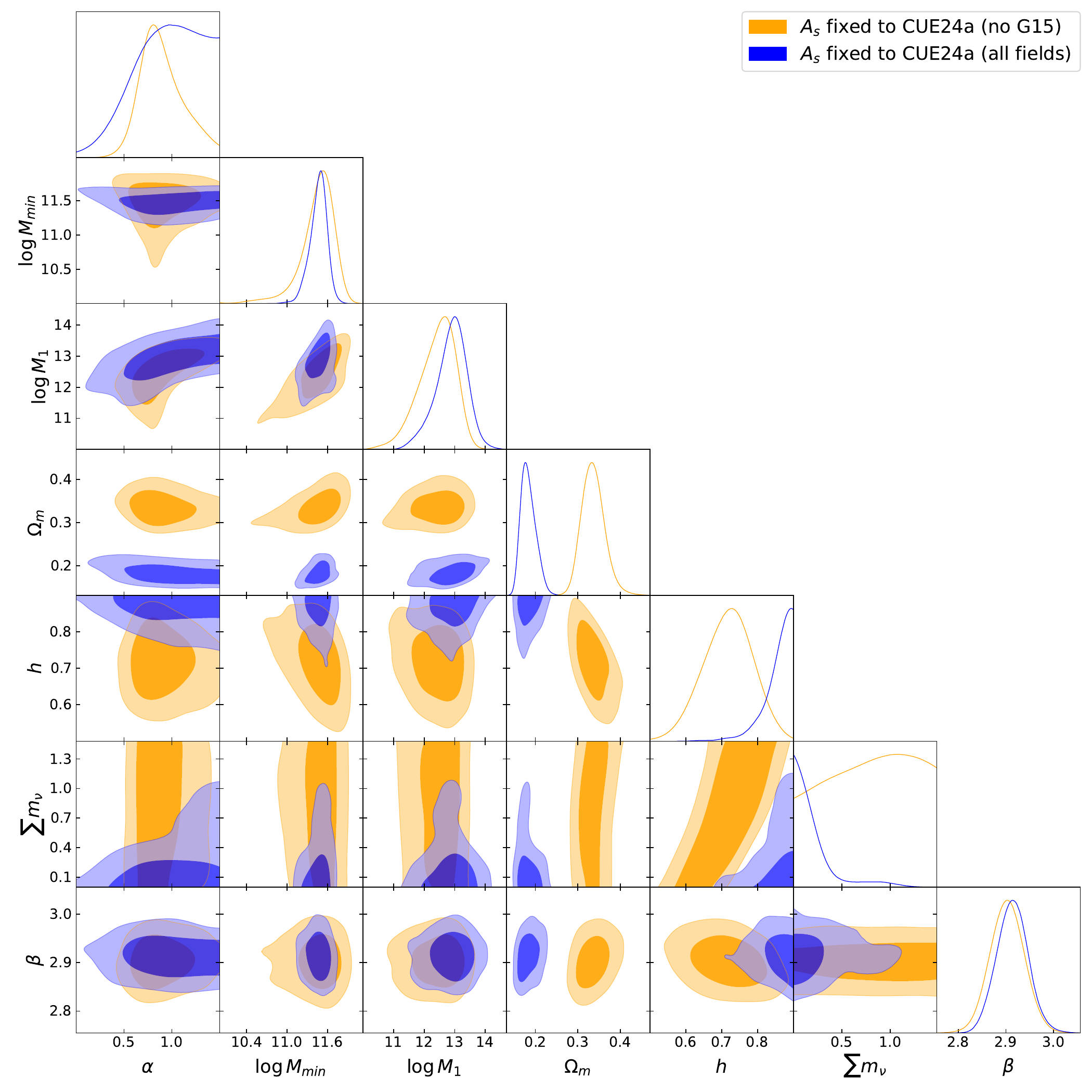}
    \caption{Marginalized posterior distributions and probability contours for a fixed $A_s$ value according to the best fit from CUE24a. The results using all four fields are shown in blue, while the case where the G15 region was excluded is depicted in orange. }
    \label{cornerplot_As_fixed_all_vs_noG15}
\end{figure}

\end{document}